\documentclass[12pt]{article}
\usepackage{amsfonts}
\usepackage{amsmath}
\usepackage{epsfig}
\usepackage{latexsym}
\usepackage{amssymb}
\usepackage{graphicx}

\def\ccA{\mathcal{A}}
\def\ccH{\mathcal{H}}
\def\ccD{\mathcal{D}}
\def\ccP{\mathcal{P}}
\def\ccV{\mathcal{V}}
\def\ccS{\mathcal{S}}
\def\gS{\mathfrak{S}}
\def\ccX{\mathcal{X}}
\def\ccF{\mathcal{F}}
\def\ccE{\mathcal{E}}
\def\ccL{\mathcal{L}}
\def\ccM{\mathcal{M}}

\def\H{\mathcal{H}}

\def\M{\mathcal{M}}
\def\B{\mathfrak{B}}
\def\T{\mathfrak{T}}
\def\bR{\mathbf{R}}
\def\bra #1{\langle #1\vert}
\def\ket #1{\vert #1\rangle}
\def\braket #1#2{\langle #1 \vert #2\rangle}
\def\ketbra #1#2{\vert #1\rangle \langle #2\vert}
\newcommand{\id}{{\rm Id}\,}
\newcommand{\Tr}{{\rm Tr}\,}

\newcounter{defin}  \newcounter{lemma}  \newcounter{theorem}
\newcounter{property} \newcounter{corol}  \newcounter{remark} \newcounter{example}

\newenvironment{lemma}{\par\refstepcounter{lemma}%\noindent
     \textbf{Lemma \thelemma.} }{\rm\par}
\newenvironment{theorem}{\par\refstepcounter{theorem}%\noindent
     \textbf{Theorem \thetheorem.}\ }{\rm\par}

\newenvironment{corollary}{\par\refstepcounter{corol}%\noindent
     \textbf{Corollary \thecorol.} }{\rm\par}
\newenvironment{definition}{\par\refstepcounter{defin}%\noindent
     \textbf{Definition \thedefin.}\ }{\rm\par}

\begin{document}

\title{CODING THEOREMS FOR HYBRID CHANNELS. II\thanks{Work partially supported by RFBR (grant No 12-01-00319) and Russian Quantum Center.}}

\author{Kuznetsova~A.\,A.\thanks{N.\,E.\,Baumann MHTU, Moscow, Russia (kuznetsova.a.a@bk.ru).} \and Holevo~A.\,S.\thanks{Steklov Mathematical Institute, RAS, Moscow, Russia (
holevo@mi.ras.ru).}}
\date{}
\maketitle
\begin{abstract}
The present work continues investigation of the capacities of measurement (quantum-classical) channels in the most general setting, initiated in~\cite{HCT}. The proof of coding theorems is given for the classical capacity and entanglement-assisted classical capacity of the measurement channel with arbitrary output alphabet, without assuming that the channel is given by a bounded operator-valued density.
\end{abstract}

\section{Introduction}\
The present work continues investigation of the capacities of measurement channels in the most general setting, initiated in~\cite{HCT}. The proof of coding theorems is given for the classical capacity  (theorem~\ref{main1}) and entanglement-assisted classical capacity (theorem~\ref{cea}) of the measurement channel with arbitrary output alphabet under the minimal regularity assumptions.  The statement of theorem~\ref{cea} was proved previously in \cite{HCT} under additional assumption that the channel is given by a bounded operator-valued density. In the present work we relax this restriction by using a generalization of the Radon-Nikodym theorem for probability operator-valued measures~\cite{H-entbrch}. The result obtained is illustrated by an example of homodyne measurement in quantum optics.

We remark that the entanglement-assisted classical capacity was studied by a number of authors under the names {\it purification capacity, measurement strength, forward classical communication cost}. In the recent paper~\cite{BRW}, where one can find further references, its alternative interpretation is developed. It is shown that a (finite-dimensional) measurement channel can be asymptotically simulated by transmission of a classical message of the size equal to the maximal entropy reduction, assisted with sufficient classical correlation between the input and the output. The result can be considered as a quantum reverse Shannon theorem in which entanglement and quantum channel are replaced, correspondingly, by classical correlation and classical channel.

\section{Preliminaries}\
Let ${\ccH}$ be a separable Hilbert space. We use the following notations: $\B(\H)$~ is the algebra of all bounded operators, $\T(\H)$~ is the space of trace-class operators in $\H$, ${\gS}({\ccH})$~ is its convex subset of {\it density operators} (i.e. positive operators with unit trace), called also quantum states.

We introduce the measure space $(\Omega, {\ccF}, \mu),$ where $\Omega$~ is a complete separable metric space, ${\ccF}$~ is a
$\sigma$-algebra of its subsets, $\mu$~ is a $\sigma$-finite measure on ${\ccF}$.
A hybrid (classical-quantum) system is described by von Neumann algebra ${\ccL} = {\ccL}^\infty(\Omega,
{\ccF}, \mu; \B(\H))$, consisting of weakly measurable, essentially bounded functions $X(\omega)$, $\omega \in \Omega$ with values in $\B(\H)$.
Consider the preadjoint space ${\ccL}_* = {\ccL}_1(\Omega, {\ccF}, \mu; \T(\H)),$ the elements of which are measurable functions $S = \{S(\omega) \}$
with values in $\T(\H),$ integrable with respect to the measure~$\mu$. An element $S = \{S(\omega) \} \in {\ccL}_*$ such that
$$
S(\omega) \ge 0\quad ({\rm mod}\ \mu),\qquad
\int_\Omega \Tr S(\omega) \,\mu(d \omega) = 1,
$$
is called {\it state} on the algebra ${\ccL}$. In notations of entropic characteristics of hybrid systems we will use the index ``cq'', of classical and quantum systems ---
the indices ``c'' and ``q'' correspondingly.

Following \cite{BL-2}, we introduce the notions of entropy and relative entropy of cq-states.
Concerning the definitions and properties of quantum entropies see e. g.~\cite{QSCI}.

\begin{definition}\label{cq-ent}\rm
The entropy of a cq-state $S$ is defined by the relation
$$
H_{\rm cq}(S) = \int_\Omega H_q(S(\omega))\,\mu(d\omega),
$$
where $H_q(S) = - \Tr S \log S$~ is the von Neumann entropy of positive operator $S \in \T(\H)$.
\end{definition}

Note that
\begin{equation} \label{cq-ent-2}
 H_{\rm cq}(S) = H_c(p) + \int_\Omega p(\omega) H_q(\widehat{S}(\omega))\,\mu(d\omega),
\end{equation}
where $p(\omega) = \Tr S(\omega),$ $\widehat{S}(\omega)=(p(\omega))^{-1}S(\omega),$ $H_c(p)$~ is the differential entropy of the probability distribution with
the density $p(\omega)$ with respect to the measure~$\mu$.

\begin{definition}\rm The {\it relative entropy} of {\rm cq}-states $S_1, S_2$
is defined by the relation
$$
H_{\rm cq}(S_1 \,\|\, S_2) = \int_\Omega H_q(
S_1(\omega)\,\|\,S_2(\omega))\,\mu(d\omega),
$$
where
$$
H_q( S_1(\omega)\,\|\,S_2(\omega))= \Tr S_1(\omega) (\log  S_1(\omega) - \log S_2(\omega))
$$
is the quantum relative entropy.
\end{definition}

To describe measurement channels we will need the following definition.
%обобщение определения наблюдаемой (см. \cite{QSCI}) для случая, когда множество исходов --- измеримое пространство $\Omega$.

\begin{definition}\rm {\it Probability operator-valued measure} $($POVM$)$ on $\Omega$  is a family $M =\{ M(A),\ A \in {\ccF}\}$
of bounded Hermitian operators in $\H,$ satisfying the conditions:

$1)$ $M(A) \ge 0,$ $A \in  {\ccF};$

$2)$ $M(\Omega) = I,$ where $I$ is the unit operator in $\H$;

$3)$ for arbitrary countable decomposition $A = \bigcup A_i$ $( A_i \cap A_j = \varnothing,\ i \neq j)$, the relation $M(A) = \sum_i M(A_i)$ holds in the
sense of weak convergence of operators.

POVM defines a {\it quantum observable} with values in $\Omega$. The {\it probability distribution} of observable~$M$ in the state~$S$ is given by the formula
\begin{equation}\label{vozm}
P_S(A)=\Tr S M(A),\qquad A \in {\ccF}.
\end{equation}
For brevity, we sometimes write $P_S(d\omega)=\Tr S M(d\omega)$.
\end{definition}

If POVM $M(d\omega)$ is defined by the density $P(\omega)$ with respect to scalar $\sigma$-finite measure $\mu,$ where $P(\omega)$~ is a uniformly bounded (with respect to the oprator norm) weakly measurable operator-valued function, then its probability distribution has the density $p_S(\omega)=\Tr S P(\omega)$ with respect to the measure~$\mu$. This case is studied in~\cite{HCT}.

In the general case the following lemma holds (a generalization of the Radon-Nikodym theorem for POVM~\cite{H-entbrch}).

\begin{lemma} \label{R-N} For an arbitrary POVM on a separable metric space ~$\Omega$ there exist a dense subspace ${\ccD} \in \H,$
a $\sigma$-finite measure $\mu$ on $\Omega,$ a countable set of Borel functions $\omega \rightarrow a_k(\omega)$ where for almost all $\omega$ the $a_k(\omega)$ are linear functionals on ${\ccD},$ satsfying the conditions
\begin{equation} \label{norm}
 \int_\Omega \sum_k |\braket {a_k(\omega)} {\psi}|^2\, \mu(d\omega) = \|\psi\|^2,\qquad \psi \in {\ccD},
 \end{equation}
\begin{equation}\label{rnt}
 \braket {\psi} {M(A)\psi} = \int_A \sum_k |\braket {a_k(\omega)} {\psi}|^2\,\mu(d\omega),\qquad \psi \in {\ccD}.
\end{equation}
\end{lemma}

In \cite{H-entbrch} it is shown that for ${\ccD}$ one can take ${\rm lin}\, \{\varphi_i\}$~---\ the linear span of a fixed orthonormal basis $\{\varphi_i\}$ .

\begin{lemma} \label{density} For arbitrary observable $M(d\omega)$ with values in~$\Omega$ and a density operator $S \in \gS(\H)$, the probability distribution $P_S (d\omega)=\Tr S M(d\omega)$ has density $p_S(\omega)$ with respect to measure $\mu$.
\end{lemma}

\textit{Proof.} Consider the spectral decomposition of the state $S$:
\begin{equation}\label{sr}
S=\sum_{i=1}^{\infty} \lambda_i \ketbra {\varphi_i} {\varphi_i}.
\end{equation}
%и последовательность операторов $S_n = \sum_{i=1}^n \lambda_i \ketbra {\varphi_i}{\varphi_i}.$
Apply lemma \ref{R-N}, with ${\ccD} = {\rm lin} \{\varphi_i\}$. For all $A \in {\ccF}$ the equality holds
\begin{equation}\label{l1}
P_S(A)\equiv\Tr S M(A) = \int_A p_S(\omega)\,\mu(d\omega),
\end{equation}
where
$$
p_S(\omega)=\sum_{i=1}^{\infty} \lambda_i  \sum_k |\braket {a_k(\omega)} {\varphi_i}|^2
$$
is a nonnegative integrable function is a nonnegative integrable function by the condition~(\ref{norm}) and the spectral decomposition~(\ref{sr}). The lemma is proved.

Let us fix an orthonormal system $\{ e_k\}$ in $\H$. According to the same conditions~(\ref{norm}) and~(\ref{sr}), the relation
\begin{equation}\label{apost}
 \widehat{S}(\omega) = (p_S(\omega))^{-1}\sum_{i=1}^{\infty} \lambda_i \sum_{j,k} \ket {e_k} \braket {a_k(\omega)}{\varphi_i}\overline{\braket {a_j(\omega)}{\varphi_i}}\bra {e_j}
\end{equation}
for $P_S$-almost all $\omega$ defines a density operator in $\H$, which we will call {\it posterior state}. The meaning of this term is that under certain conditions the operator
$\widehat{S}(\omega)$ describes state of the quantum system after measurement of observable~$M$, which resulted with the outcome~$\omega$~\cite{QSCI}.

Following \cite{Sh-ERQM}, define the {\it entropy reduction} by the relation
\begin{equation}\label{er-1}
{\rm ER}\,(S, M) = H_q(S) -  \int_\Omega p(\omega)H_q(\widehat{S}(\omega))\,\mu(d\omega),
\end{equation}
which is consistent provided $H_q(S)<\infty$. We mention the following approximation properties.
Consider a sequence of states $S_n = \sum_{i=1}^n \widetilde{\lambda_i}\ketbra {\varphi_i} {\varphi_i},$ where $\widetilde{\lambda_i} = (\sum_{k=1}^n \lambda_k)^{-1} \lambda_i.$
By lemma~4 of the paper~\cite{L}, the above sequence $S_n$ satisfies the condition
\begin{equation} \label{cond1}
 \lim_{n \rightarrow \infty}H_q(S_n)=H_q(S)<\infty.
 \end{equation}
According to the theorem 2 from  \cite{Sh-ERQM}, this implies
  \begin{equation} \label{erconv}
\lim_{n \rightarrow \infty} {\rm ER}\,(S_n,{M}) = {\rm ER}\,(S,{M})<\infty.
\end{equation}

\section {The classical capacity of a measurement channel}\quad\\

\vspace{-8pt}
\begin{definition} \label{M-ch}\rm
Let $M$ be a POVM, $P_S$~---\ its probability distribution in the state~$S$, which is given by the formula~{\rm(\ref{vozm})}. {\it Measurement channel} $\M$ is an affine map $S\rightarrow P_S(d\omega)$ of the convex set
of quantum states ${\gS}(\H)$ into the set of probability distributions on~$\Omega$.
\end{definition}

To apply the method of block coding, we need to define the $n$-th
degree $\M^{\otimes n}$ of the channel $\M$. Let $\H^{\otimes n}$ be
the $n$-th tensor degree of the Hilbert space $\H$ and let
$(\Omega^{\times n}, {\ccF}^{\times n})$ be the product of $n$
copies of the measurable space $(\Omega, {\ccF})$. The cnannel
$\M^{\otimes n}$ is defined by the observable $M^{\otimes n}$ with
values in $\Omega^{\times n}$ such that
$$
M^{\otimes n}(A_1\times\cdots\times A_n)=M(A_1)\otimes\cdots\otimes M(A_n).
$$
By using an analog of the extension theorem for POVM, one can show
that this relation defines uniquely all the values $M^{\otimes
n}(A^{(n)})$, $A^{(n)}\in {\ccF}^{\times n}.$

In the case of infinite-dimensional $\H$ one usually introduces a
constraint onto the input states of the channel (otherwise the
capacities are infinite as a rule). Let $F$ be a positive
selfadjoint (in general unbounded) operator in the space $\H$, with
the spectral decomposition $F = \int_0^{\infty} x \,d E(x),$ where
$E(x)$ is the spectral function. We introduce the subset of states
 \begin{equation} \label{setA}
{\ccA}_E=\{S\in\gS(\H):  \Tr SF \le E\},
\end{equation}
where $E$ is a positive constant, and the trace in~(\ref{setA}) is
understood as the integral $\int_0^{\infty} x \,d (\Tr S E(x))$ (for
more detail see~\cite{ClCap}). Notice that if the operator $F$
satisfies the condition
\begin{equation} \label{condF}
\Tr \exp (-\beta F) < \infty\quad \mbox{для всех}\ \beta > 0,
\end{equation}
then $ H_q(S) < \infty$ for all $S$ such that $ \Tr SF \le E$
(see~\cite{ClCap}). The corresponding constraint for the channel
$\M^{\otimes n}$ is determined by the operator
$$
F^{(n)} = F \otimes I \otimes \cdots \otimes I + \cdots + I \otimes I \otimes \cdots \otimes F.
$$
Denote
\begin{equation} \label{Fn}
{\ccA}^{(n)}_E=\{S^{(n)}\in\gS(\H^{\otimes n}): \Tr S^{(n)}F^{(n)}  \le nE\}.
\end{equation}

\begin{definition}\label{code}\rm The {\it code} of length $n$ and size $N$
is a pair $(\Sigma^{(n)}, {\ccV}^{(n)})$, where:

{\rm 1)} $\Sigma^{(n)} = \{S_i^{(n)},\ i = 1,\ldots,N \}$ is a
family of states from ${\ccA}_E^{(n)}$;

{\rm 2)} ${\ccV}^{(n)}= \{V_j,\ j=0, 1,\ldots,N\}$ is a
decomposition of the space $\Omega^{\times n}$.
\end{definition}

\begin{definition}\rm The {\it average error probability} of the code $(\Sigma^{(n)}, {\ccV}^{(n)})$
is the quantity
\begin{equation}
\overline u (\Sigma^{(n)}, {\ccV}^{(n)}) = \frac 1 N \sum_{j=1}^N \big( 1 -
 \Tr S_j^{(n)}M^{\otimes n}(V_j) \big).
\end{equation}
 \end{definition}

We denote by $u(n,N)$ the greatest lower bound of the quantity
$\overline{u}(\Sigma^{(n)}, {\ccV}^{(n)})$ with respect to all codes
of length $n$ and size $N$.

\begin{definition}\rm We call the {\it classical capacity} $C({\ccM}, {\ccA}_E)$ of the measurement channel
${\ccM}$ with the constraint (\ref{Fn}) the supremum of all
achievable rates i.e.\ the values $R>0$, satisfying the condition
\begin{equation}
\lim_{n \rightarrow \infty} u(n, 2^{nR}) = 0.
\end{equation}
\end{definition}

We call by {\it ensemble} of states a finite probability
distribution $\pi=\{\pi_x; S_x\}$ on the set of states
${\gS}({\ccH})$, ascribing probabilities $\pi_x$ to certain states
$S_x$. The {\it average state} of ensemble is defined as:
$\overline{S}_\pi = \sum_x \pi_x S_x$. Let us denote ${\ccP}_E$ the
set of ensembles $\pi$ such that $\overline{S}_\pi \in {\ccA}_E$;
similarly, we denote ${\ccP}_E^{(n)}$ the set of ensembles
$\pi^{(n)}$ in $\gS(\H^{\otimes n})$, the average state of which
satisfying the condition~(\ref{Fn}).

For given measurement channel ${\ccM}$ and ensemble $\pi$ define the
quantity
\begin{equation}\label{chi-c}
 I (\pi, {\ccM}) =
 \sum_x \pi_x \int_\Omega p(\omega|x) \log
\frac {p(\omega|x)} {\overline{p}(\omega)} \, \mu(d\omega).
\end{equation}
Here $\overline{p}(\omega)$ and $p(\omega | x)$~are the probability
densities of the distributions $\overline{P}(d\omega)=\Tr
\overline{S}_\pi M(d\omega)$ and $P_x(d\omega)=\Tr S_x M(d\omega)$
correspondingly. The quantity $I (\pi,{\ccM})$ is the Shannon mutual
information between the discrete random variable $X$, having the
probability distribution $\{\pi_x\}$ and the random variable
$\omega$ with conditional probability density $p(\omega\,|\,x)$,
defined via lemma~\ref{density}. Notice that there is a
representation of the quantity $I (\pi, {\ccM})$ as a supremum over
decompositions ${\ccV} = \{ V_i \}$ of the output space $\Omega$
(cf.~\cite[formula~(1.2.3)]{D}):
\begin{equation}\label{main2}
I (\pi, {\ccM}) =  \sup_{{\ccV}}  \bigg( \sum_i \sum_x \pi_x  P_x(V_i) \log
\frac{P_x (V_i)}{\overline{P} (V_i)} \bigg).
\end{equation}
The quantity under supremum is equal to $I (\pi, {\ccM}_{\ccV})$,
where ${\ccM}_{\ccV}$~is the measurement channel, corresponding to
the discrete observable $\{M(V_i)\}$.

\begin{theorem}\label{main1}
The classical capacity of the measurement channel $\M$ with the
constraint~{\rm(\ref{Fn})} is given by the relation
\begin{equation}
 C(\ccM, {\ccA}_E) = \sup_{\pi \in {\ccP}_E}I (\pi, {\ccM}).
 \end{equation}
 \end{theorem}

\textit{Proof.} Denote
$$
C_n=\sup_{\pi^{(n)} \in {\ccP}^{(n)}_E}I (\pi^{(n)}, {\ccM}^{\otimes n}).
$$
We need to show that
$$
 C(\ccM, {\ccA}_E) =C_1.
$$
Let us first establish the additivity property $C_n=nC_1$.

For a fixed decomposition ${\ccV}=\{V_i\}$, the measurement channel
${\ccM}_{\ccV}$ is embedded into quantum entanglement-breaking
channel (see e.g.~\cite{H-QO}), therefore according to~[3] its
capacity is given by the expression
\begin{equation}\label{cap1}
C({\ccM}_{\ccV},{\ccA}_E)= \sup_{\pi \in {\ccP}_E}I (\pi, {\ccM}_{\ccV}),
\end{equation}
and has the additivity property (see\ [6])
$$
C({\ccM}^{\otimes n}_{\ccV},{\ccA}^{(n)}_E)=nC({\ccM}_{\ccV},{\ccA}_E).
$$
Notice that similarly to (\ref{cap1}), the left-hand side is equal
to $\sup_{\pi^{(n)} \in {\ccP}^{(n)}_E}I (\pi^{(n)}, {\ccM}^{\otimes
n}_{\ccV})$, so that
\begin{equation}\label{add2}
\sup_{\pi^{(n)} \in {\ccP}^{(n)}_E}I (\pi^{(n)}, {\ccM}^{\otimes n}_{\ccV})=n\sup_{\pi \in {\ccP}_E}I (\pi, {\ccM}_{\ccV}).
\end{equation}
By using a result of R.\,L.\,Dobrushin (theorem~2.2 in~\cite{D}), we
have
$$
I (\pi^{(n)}, {\ccM}^{\otimes n})=\sup_{{\ccV}}I (\pi^{(n)}, {\ccM}^{\otimes n}_{\ccV}),
$$
because the supremum in the right-hand side is equal to the supremum
of the information with respect to decompositions of the space
$\Omega^{\times n}$ of special form, consisting of products
$V_1\times\cdots\times V_n$ of the sets from the
decomposition~${\ccV}$. The class of all such products has the
ordering property that is required for validity of theorem~2.2
in~\cite{D}. Hence
\begin{eqnarray*}
C_n&=&\sup_{\pi^{(n)} \in {\ccP}^{(n)}_E}\sup_{{\ccV}}I (\pi^{(n)}, {\ccM}^{\otimes n}_{\ccV})=\sup_{{\ccV}}\sup_{\pi^{(n)} \in {\ccP}^{(n)}_E}I (\pi^{(n)}, {\ccM}^{\otimes n}_{\ccV})\\
&=&n \sup_{{\ccV}}\sup_{\pi \in {\ccP}_E}I (\pi, {\ccM}_{\ccV})
=n \sup_{\pi \in {\ccP}_E}\sup_{{\ccV}}I (\pi, {\ccM}_{\ccV})\\
&=&n\sup_{\pi \in {\ccP}_E}I (\pi, {\ccM})=nC_1,
\end{eqnarray*}
where we used (\ref{add2}) in the third equality.

Now let us prove the inequality $C({\ccM},{\ccA}_E) \le C_1.$
Without loss of generality we can suppose that $C_1<\infty.$ Let $R
> C_1.$ By applying Fano's inequality, we obtain similarly to the
relation~(10.19) in~\cite{QSCI}
$$
u(n,2^{nR})\ge 1-\frac{C_n}{nR}-\frac{1}{nR}= 1-\frac{C_1}{R}-\frac{1}{nR},
$$
where in the second equality we used the additivity $C_n=nC_1$.
Therefore $\liminf_{n \rightarrow \infty}u(n,2^{nR})>0$, and hence
$C({\ccM},{\ccA}_E) \le C_1.$

For the proof of the converse inequality we note that
$$
C({\ccM},{\ccA}_E) \ge C({\ccM}_{{\ccV}},{\ccA}_E)=\sup_{\pi \in {\ccP}_E}I (\pi, {\ccM}_{\ccV})
$$
for arbitrary decomposition ${\ccV}$. By taking supremum over the
decompositions ${\ccV}$, we obtain $C({\ccM},{\ccA}_E) \ge \sup_{\pi
\in {\ccP}_E}I (\pi, {\ccM})=C_1$. The theorem~1 is proved.

\section{Entanglement-assisted capacity of a measurement channel}\
Consider the following protocol of classical information transmission through the measurement channel $\M$.
Transmitter $A$ and receiver $B$ are in the pure entangled state
состоянии $S_{AB} = \ketbra {\psi} {\psi}$, where  $\ket {\psi} = \sum_j c_j \ket {e_j} \otimes \ket
{\widetilde{e_j}},$ % $\{\ket {e_i} \otimes \ket {\widetilde{e_j}} \}$ ---
%ортонормированный базис в пространстве $\H_A \otimes \H_B$,
satisfying the condition $H_q(S_A) = H_q(S_B) < \infty.$

Let ${\ccX}$~be a finite alphabet, and the classical signal $x \in {\ccX}$ appears with probability $\pi_x$. The party $A$ performs encoding $x \rightarrow {\ccE}_x$ and sends its part of the resulting common state via the channel $\M$. Thus the party $B$ has at its disposal the hybrid system $\Omega B$, where $\Omega $~is the classical system at the output of the measurement channel. After the measurement of observable $M(d\omega)$, the state in the hybrid system is described in the following way:
$$
\sigma_x(d\omega) = \sum_{j,k} c_j \overline{c}_k [\Tr
{\ccE}_x (\ketbra {e_j} {e_k}) M(d\omega)] \ketbra {\widetilde e_j}
{\widetilde e_k}.
$$
Then the party $B$ may perform measurement of an observable in the system $\Omega B$, extracting in this way information about the signal~$x$.

With the block coding, the encoded states transmitted through the
channel $\M^{\otimes n} \otimes \id_B^{\otimes n}$, have the form
\begin{equation} \label{ac}
S_{\alpha}^{(n)} = ({\ccE}_\alpha^{(n)} \otimes \id_B^{\otimes n}) [S_{AB}^{(n)}],
\end{equation}
where $S_{AB}^{(n)}$ is pure entangled state for $n$ copies of the
system $AB$, satisfying the condition $H(S_B^{(n)}) < \infty,$
$\alpha$ is the classical message (e.g. a word in an alphabet
${\ccX}$), $\alpha \rightarrow {\ccE}_\alpha^{(n)}$~are the
encodings for $n$ copies of the system~$A$. The input states of the channel
$\M^{\otimes n}$ are subject to the constraint~(\ref{Fn}), which is equivalent to similar
constraint for the channel $\M^{\otimes n}
\otimes \id_B^{\otimes n}$ with the operators $F^{(n)} \otimes
I_B^{\otimes n}.$

For the channel $\M$ with the input constraint (\ref{setA}) we consider the quantity
\begin{equation}\label{Cea-cq-n}
C_{\rm ea}^{(n)}(\M^{\otimes n}, {\ccA}_E^{(n)}) =  \sup_{(\pi^{(n)}_{\alpha}, S^{(n)}_{\alpha})} \chi_{\rm cq} \left( \{\pi_\alpha^{(n)}\}; \{(\M^{\otimes n} \otimes \id_B^{\otimes n})S^{(n)}_\alpha\} \right) ,
\end{equation}
where
$$
\chi_{\rm cq} \left( \{\pi_x\}; \{S_x\} \right) =  H_{\rm cq}\bigg(\sum_x \pi_x S_x\bigg) - \sum_x
\pi_x H_{\rm cq}(S_x),\qquad S_x \in {\ccL}_*,
$$
and the supremum is taken over all state ensembles of the form (\ref{ac}), satisfying the condition
$$
\sum_\alpha \pi^{(n)}_\alpha \Tr S^{(n)}_\alpha (F^{(n)} \otimes I_B^{\otimes n}) \le nE.
$$

The {\it classical entanglement-assisted capacity} for the quantum-classical channel $\M$ with the constraint~(\ref{setA}) is defined by the relation
$$
C_{\rm ea}(\M, {\ccA}_E) = \lim_{n \rightarrow \infty} \frac 1 n\,
C_{\rm ea}^{(n)}(\M^{\otimes n},{\ccA}_E^{(n)}).
$$

\begin{theorem} \label{cea} Let $\M$~be an arbitrary measurement channel with the input constraint~{\rm(\ref{Fn})}.
Assume,  that the operator $F$ satisfies the condition~{\rm(\ref{condF}),} and the channel~$\M$ satisfies the condition
\begin{equation} \label{condM}
 \sup_{S_A:  \Tr S_A F \le E} H_c(p_{S_A}) < \infty,
\end{equation}
where $H_c(p_{S_A})$~is the classical differential entropy of the probability density of the output
distribution of the channel~$\M$.
Then the entanglement-assisted capacity is given by the expression
\begin{equation}\label{cer}
C_{\rm ea}(\M, {\ccA}_E) = \sup_{S_A:  \Tr S_A F \le E} {\rm ER}\,(S_A, M).
\end{equation}
\end{theorem}

\textit{Proof.} %Сначала докажем неравенство \le в формуле \ref{cer}, т.е.
%\begin{equation}\label{neq}
%C_{ea}(\M, {\ccA}_E) \le \sup_{S_A:~\Tr S_A F \le E} ER(S_A, M),
%\end{equation}
%для чего покажем, что $C_{ea}^{(1)}(\M, {\ccA}_E) \le
%\sup_{S_A:~\Tr S_A F \le E} ER(S_A, M).$
In the proof we use the corresponding result for measurement channels defined by a bounded operator density, obtained in~\cite{HCT}.

Let $S_{AB}$ be the initial entangled state of the system~$AB$.
After applying encoding ${\ccE}_A^x$ in the system~$A$ the state of the composite system is described by the operator
$$
S_{AB}^x = ({\ccE}_A^x \otimes \id_B)S_{AB}
$$
with the partial states $S_A^x = {\ccE}_A^x (S_A)$ and $S_B^x = S_B$.

To establish the inequality $\le$ in the formula (\ref{cer}), it is sufficient to prove (see \cite{HCT} for detail) that
\begin{equation} \label{eq1}
H_{\rm cq}\bigg( \sum_x \pi_x (\M \otimes \id_B)S^x_{AB}\bigg) - \sum_x \pi_x H_{\rm cq} \left(\M \otimes \id_B(S_{AB}^x)\right) \le {\rm ER}\,(\overline{S}_A, M).
\end{equation}
Here $\overline{S}_A = \sum_x \pi_x S_A^x$, and the constraint~(\ref{Fn}) implies the condition
\begin{equation}\label{ogr}
\Tr \overline{S}_A F \le E.
\end{equation}
A result of \cite{HCT} implies that the relation  (\ref{eq1}) holds for finite-rank states $\overline{S}_A$ satisfying the constraint (\ref{ogr}). For the proof in the general case
we apply approximation by finite-rank states.

Assume first that $\Tr \overline{S}_A F \le E' < E$ for a positive $E'.$ Let $\overline{S}_A$ have the spectral decomposition $\overline{S}_A = \sum_i \lambda_i \ketbra {\varphi_i} {\varphi_i}$.
Consider the increasing sequence of projections  $P_n=\sum_{i=1}^n \ketbra {\varphi_i} {\varphi_i}$ converging to the unit operator $I_A$, and the sequence of states
$$
S_{AB}^x(n) =P_n \otimes I_B S_{AB}^x P_n \otimes I_B +\ketbra {\phi} {\phi} \otimes (S_B - S_B^x(n)),
$$
where $ S_B^x(n) = \Tr_A (P_n \otimes I_B S_{AB}^x P_n \otimes I_B), $  $\ket {\phi}$~is a fixed unit vector from ${\rm lin} \{\varphi_i\}$, belonging to the domain of $\sqrt{F}$. The partial states of $S_{AB}^x(n)$ are
$$
\Tr_B S_{AB}^x(n) = S_A^x(n)=P_n S_A^x P_n +(1-\Tr P_n S_A^x)\ketbra {\phi} {\phi},\quad\Tr_A S_{AB}^x(n) = S_B.
 $$
Then the average state in the system $A$ is equal to
 $$
 \overline{S}_A(n) = \sum_x \pi_x S_A^x(n) = P_n \overline{S}_A P_n + (1-\Tr P_n \overline{S}_A)\ketbra {\phi} {\phi}.
$$
We have $\overline{S}_A(n) = \sum_{i=1}^n {\lambda_i} \ketbra {\varphi_i} {\varphi_i} + (1-\Tr P_n \overline{S}_A)\ketbra {\phi} {\phi},$ then $\|\overline{S}_A(n) - \overline{S}_A\|_1\rightarrow 0$ for $n \rightarrow\infty$ and
$$
\Tr \overline{S}_A(n) F = \sum_{i=1}^n {\lambda}_i
\|\sqrt{F} \varphi_i\|^2 + (1-\Tr P_n \overline{S}_A) \braket {\phi} {F \phi} \le E'+\varepsilon_n,
$$ где $\varepsilon_n \rightarrow 0$ при $n
\rightarrow \infty.$
Thus, starting from some value of $n$, the density operator $\overline{S}_A(n)$ satisfies the input constraint~(\ref{ogr}).

Using the condition (\ref{condM}), similarly to the proof of the coding theorem for measurement of observable in~\cite{HCT} we obtain the inequality~(\ref{eq1}) for the ensemble $\{ \pi_x, S^x_{AB}(n)\},$ which can be written in the following form based on the relative entropy:
\begin{equation} \label{eq3}
\sum_x {\pi}_x H_{\rm cq}\bigg((\M \otimes \id_B)(S^x_{AB}(n))\,\bigg\|\,\sum_x {\pi}_x (\M \otimes \id_B)S^x_{AB}(n)\bigg) \le {\rm ER}\,( \Tr \overline{S}_A(n), M).
\end{equation}
Take the limit $n \rightarrow \infty$ in (\ref{eq3}). By noting that $\lim_{n \rightarrow \infty} H_q(\overline{S}_A(n))= H_{q}(\overline{S}_A)$, using theorem 2 from \cite{Sh-ERQM} (i.e.\ the equality~(\ref{erconv})) in the left-had side and the lower semicontinuity of the relative entropy in the right-hand side, we obtain~(\ref{eq1}) for the ensemble $\{ \pi_x, S^x_{AB}\}$.

Now consider the case  $\Tr \overline{S}_A F = E$. Take a unit vector $\ket {e} \in {\rm lin}\{\phi_i\},$ satisfying the condition $\braket {e} {Fe} < E,$ and construct the approximation $S^x_{AB}(\varepsilon) = (1-\varepsilon) S^x_{AB} + \varepsilon \ketbra {e} {e} \otimes S_B, ~ 0 < \varepsilon < 1.$ Then the average state of the system $A$
is $\overline{S}_A({\varepsilon}) = (1-\varepsilon)\overline{S}_A + \varepsilon \ketbra {e} {e}$,  and the following condition holds
$$
 \Tr \overline{S}_A({\varepsilon}) F < E.
 $$
Let us repeat previous argument approximating $S^x_{AB}(\varepsilon)$  by the states of the form
$$
(1-\varepsilon) P_n \otimes I_B  S^x_{AB} P_n \otimes I_B + \ketbra e e (S_B - (1-\varepsilon) \Tr_A P_n \otimes I_B S^x_{AB})
$$
with the partial states $(1-\varepsilon)S_A^x +  (1 - (1-\varepsilon) \Tr P_n S_A^x) \ketbra e e$ and $S_B$ in the systems $A$ and $B$ correspondingly. We obtain that the inequality
(\ref{eq1}) holds for $S^x_{AB}(\varepsilon)$, $\overline{S}_A({\varepsilon})$.
Since $\lim_{\varepsilon \rightarrow 0} H_q (\overline{S}_A({\varepsilon})) = H_q(\overline{S}_A),$  then, %по теореме о промежуточной последовательности, так как для энтропии $\overline{S}_A({\varepsilon})$ %справедлива следующая оценка (см. \cite{W})
%$$
%(1-\varepsilon)H_q(\overline{S}_A) + \varepsilon H_q( \ketbra {e} {e}) \le H_q (\overline{S}_A({\varepsilon})) \le (1-\varepsilon)H_q(\overline{S}_A) + \varepsilon H_q( \ketbra {e} {e}) - %(1-\varepsilon)\log (1-\varepsilon) - \varepsilon \log \varepsilon,
%$$
tending $\varepsilon$ to zero we obtain~(\ref{eq1}) for ensembles satisfying the condition $\Tr \overline{S}_A F = E$. The rest of the proof is similar to the case of observable with a bounded density~\cite{HCT}.

To prove the inequality $\ge$ in (\ref{cer}) we consider an arbitrary state $S \in \gS(\H)$, $S = \sum_{i=1}^\infty \lambda_i \ketbra {\varphi_i} {\varphi_i},$ satisfying the input constraint. Apply lemma 1, setting ${\ccD} = {\rm lin}\,\{\varphi_i\}$ and defining posterior states $\widehat{S}(\omega)$ by the relation~(\ref{apost}).
Then the argument is similar to trhe proof of proposition~4 from~\cite{ClCap}, and also theorem~3 from~\cite{HCT}. Theorem~2 is proved.

Of special interest is the case of {\it pure} POVM for which there exists a representation (\ref{rnt}) of the form
\begin{equation}\label{rntpure}
 \braket {\psi} {M(A)\psi} = \int_A |\braket {a(\omega)} {\psi}|^2\,\mu(d\omega),\qquad \psi \in {\ccD}.
\end{equation}
In this case the posterior state (\ref{apost}) is a pure state, not depending on $x$:
\begin{equation}\label{apostpure}
 \widehat{S}(\omega) = \ketbra {e}{e},
\end{equation}
where $e\in {\ccH}$~is a unit vector. Thus, $H_q(\widehat{S}(\omega))=0$ and the entropy reduction is equal to
\begin{equation}\label{er-1pure}
{\rm ER}\,(S, M) = H_q(S).
\end{equation}
The relation (\ref{cer}) takes the form
\begin{equation}\label{cerpure}
C_{\rm ea}(\M, {\ccA}_E) = \sup_{S_A: \Tr S_A F \le E} H_q(S).
\end{equation}
It is well known that this supremum is attained on the Gibbs state
\begin{equation} \label{gibbs}
S_{\beta}=c(\beta)^{-1} \exp (-\beta F), \qquad c(\beta)=\Tr \exp (-\beta F),
\end{equation}
where $\beta$ is found from the condition $\Tr S_{\beta}F=E$, and it is equal to $\beta E + c(\beta)$. Thus theorem~\ref{cea} implies the following statement.

\begin{corollary} For arbitrary measurement channel,  corresponding to pure POVM,
$$
C_{\rm ea}(\M, {\ccA}_E)=\beta E + c(\beta),
$$
where $\beta$ is found from the condition $\Tr S_{\beta}F=E$.
\end{corollary}

Let us illustrate this result by two examples. Let
${\ccH}=L^2(\bR)$, $Q$~be the operator of multiplication by $x$,
$P=-id/dx$ with the common essential domain ${\ccD}={\ccS}(\bR)$
(the space of infinitely differentiable functions rapidly decreasing
with all derivatives, see e.g.~\cite{QSCI}). The spectral measure
$M$ of the selfadjoint operator $Q$ can be represented in the form
~(\ref{rntpure}):
$$
 \braket {\psi} {M(A)\psi} = \int_A |\braket {x} {\psi}|^2 \,dx,\qquad \psi \in {\ccS}({\bR}),
$$
where $\braket {x} {\psi}=\psi (x)$, $\psi \in {\ccS}(\bR)$, are the
Dirac's $\delta$-functionals. Thus the POVM $M$ does not have
bounded operator density, the result of the paper~\cite{HCT} is not
applicable and one should apply the approach of the present paper.
Arbitrary density operator $S$ in $L^2(\bR)$ is defined by the
kernel which is conveniently written in the symbolic form
$\bra{x}S\ket{y}$ (for continuous kernels this notation can be
understood literally). Consider the channel corresponding to the
measurement of observable~$Q$, which maps a density operator~$S$
into the probability density $\bra{x}S\ket{x}$ with respect to the
Lebesgue measure on the real line. In quantum optics such a channels
describes statistics of {\it homodyne measurement} of one mode $Q,
P$ of electromagnetic field~\cite{cd}. As a constraint operator one
usually takes the oscillator energy $F=(P^2+Q^2)/2$. Notice that the
condition~(\ref{condM}) is fulfilled, as the inequality $\Tr SF \le
E$ implies
$$
\int x^2p_{S}(x)\,dx=\Tr S Q^2\le 2\Tr SF\le 2E,
$$
and the maximal differential entropy (equal to $(1/2)\log (2\pi e
(2E))$) under this constraint is attained on the Gaussian
probability density. Substituting this value of supremum, equal to
the entropy of the Gibbs state of oscillator with the mean energy
$E$ (see e.g.~\cite{cd},~\cite{QSCI}) into (\ref{cerpure}), we
obtain
\begin{equation}\label{ceahom}
C_{\rm ea}(\M, {\ccA}_E)=\bigg(E+\frac{1}{2}\bigg)\log \bigg(E+\frac{1}{2}\bigg)-\bigg(E-\frac{1}{2}\bigg)\log \bigg(E-\frac{1}{2}\bigg).
\end{equation}
On the other hand, the classical capacity of homodyne channel
computed in~\cite{cd},~\cite{hall} is equal to
\begin{equation}
C(\M_{\rm hom}, {\ccA}_E)=\log (2E).
\end{equation}

\begin{figure}[t]
\begin{center}
\includegraphics[width=207 pt, height= 185 pt]{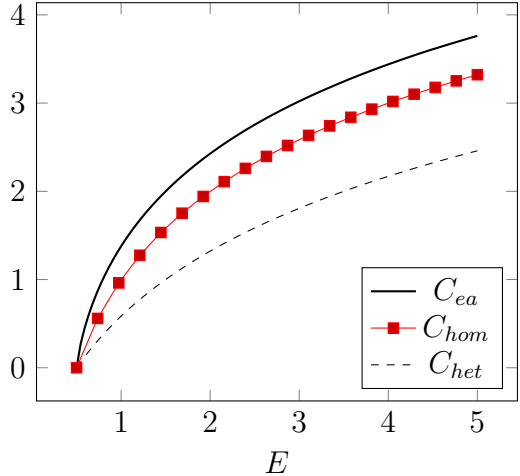}
\caption{Classical capacities of the optical measurement channels}\label{fig1}
\end{center}
\end{figure}

According to the corollary 1, the relation~(\ref{ceahom}) holds for
arbitrary pure measurement channel including {\it heterodyne
channel}, which maps a density operator $S$ into probability density
$\bra{x,y}S\ket{x,y}$ with respect to the Lebesgue measure on the
plane, where $\ket{x,y}$~are the coherent states of the quantum
oscillator~\cite{H-QO}. Notice that in this case the bounded
operator density exists and results of paper~\cite{HCT} are
applicable. The classical capacity of the heterodyne channel
computed in ~\cite{hall},~\cite{H-QO} is equal to
\begin{equation}
C(\M_{\rm het}, {\ccA}_E)=\log \bigg(E+\frac{1}{2}\bigg).
\end{equation}
For all $E>1/2$ the inequalities hold
$$
C(\M_{\rm het}, {\ccA}_E)<C(\M_{\rm hom}, {\ccA}_E)<C_{\rm ea}(\M, {\ccA}_E).
$$
The graphs of the three capacities are shown on Fig.~1.

\end{document}